\documentclass[preprint,aps,prc,showpacs,nofootinbib]{revtex4}

\usepackage{epsfig}
\usepackage{graphicx}

\begin{document}

\title{On the intrinsic limitation of the Rosenbluth method at large $Q^2$.}

\author{E. Tomasi-Gustafsson}
\affiliation{\it DAPNIA/SPhN, CEA/Saclay, 91191 Gif-sur-Yvette Cedex,
France }
\email{etomasi@cea.fr}

\date{\today}

\begin{abstract}
Correlations in the elastic electron proton scattering data show that the Rosenbluth method is not reliable for the extraction of the electric proton form factors at large momentum transfer, where the magnetic term dominates, due to the size and the $\epsilon$ dependence of the radiative corrections.
\end{abstract}

\maketitle

The determination of the elastic proton electromagnetic form factors (FFs) at large momentum transfer is a very actual problem, due to the availability of electron beams in the GeV range, with high intensity and high polarization, large acceptance spectrometers, hadron polarized targets, hadron polarimeters. The possibility to extend the measurements of such fundamental quantities, which contain dynamical information on the nucleon structure,  inspired experimental programs at JLab, Frascati and at future machines, as GSI, both in the space-like  and in the time-like regions.

The traditional way to measure electromagnetic proton form factors consists in the determination of the $\epsilon$ dependence of the reduced elastic differential cross section, which can be written, assuming that the interaction occurs through the exchange of one-photon, as \cite{Ro50}:
\begin{equation}
\sigma_{red}=\epsilon(1+\tau)\left [1+2\displaystyle\frac{E}{m}\sin^2(\theta/2)\right ]\displaystyle\frac
{4 E^2\sin^4(\theta/2)}{\alpha^2\cos^2(\theta/2)}\displaystyle\frac{d\sigma}{d\Omega}=\tau G_{Mp}^2+\epsilon G_{Ep}^2,
\label{eq:sigma}
\end{equation}
$$
\epsilon=[1+2(1+\tau)\tan^2(\theta/2)]^{-1}, 
$$
where $\alpha=1/137$, $\tau=Q^2/(4m^2)$, $Q^2$ is the momentum transfer squared, $m$ is the proton mass, $E$ and $\theta$ are the incident electron energy and the scattering angle of the outgoing electron, respectively, $G_{Mp}$ and $ G_{Ep}$ are the magnetic and the electric proton FFs.  Measurements of the elastic differential cross section at different angles, for a fixed value of $Q^2$, allow to determine $G_{Ep}$ as the slope and $G_{Mp}$ as the intercept from the linear $\epsilon$ dependence (\ref{eq:sigma}).

High precision data on the ratio of the electric to magnetic proton FFs at large $Q^2$, have been recently obtained  \cite{Jo00,Ga02}, through the polarization transfer method \cite{Re68}.  Such data revealed a surprising trend, which deviates from the expected scaling behavior previously obtained through the measurement of elastic  $ep$ cross section according the Rosenbluth separation method  \cite{An94}. New precise measurements of unpolarized elastic $ep$ cross section\cite{Ar04} and re-analysis of the old data \cite{Ch04,Ar04b} confirm that the behaviour of the measured ratio $R(Q^2)=\mu  G_{Ep}(Q^2)/G_{Mp}(Q^2)$ ($\mu=2.79$ is the magnetic moment of the proton) is different depending to the method used: 
\begin{itemize}
\item {\bf scaling behavior for unpolarized cross section measurements:} $R(Q^2)\simeq 1$; $G_{Mp}$ has been extracted up to $Q^2\simeq 31$ GeV$^2$ \cite{Ar75} and fall with $Q^2$ according to a dipole form: $G_D(Q^2)=(1+Q^2/0.71\mbox{~GeV}^2)^{-2}$;
\item  {\bf strong monotonical decreasing from polarization transfer measurements.} 
\begin{equation}
R(Q^2)==1-0.13(Q^2~[\mbox{GeV}^2]-0.04). 
\label{eq:brash}
\end{equation}
\end{itemize}
The ratio deviates from unity, as $Q^2$ increasing, reaching a value of $\simeq$ 0.34 at $Q^2\simeq $ 5.5 GeV$^2$ \cite{Br03}. 

This puzzle has given rise to many speculations and different interpretations \cite{Bl03,Gu03,Ch04b}, suggesting expensive experiments. In particular, it has been suggested that the $2\gamma  $ exchange could solve this discrepancy through its interference with the the main mechanism (the $1\gamma$ exchange). In a previous paper \cite{ETG} it has been shown that the present data do not show any evidence of the presence of the $2\gamma$ mechanism, in the limit of the experimental errors. The main reason is that, if one takes into account $C$-invariance and crossing symmetry, the $2\gamma$ mechanism introduces a non linear, very specific $\epsilon$ dependence of the reduced cross section \cite{Re1,Re03t,Re04a}, whereas the data does not show any deviation from linearity.

Before analyzing the data in a different perspective, we stress the following points:
\begin{itemize}
\item No experimental bias has been found in both types of measurements, the experimental observables being the differential cross section on one side, and the polarization of the outgoing proton in the scattering plane (more precisely the ratio between the longitudinal and the transverse polarization), on the other side.
\item The discrepancy is not at the level of these observables: it has been shown that constraining the ratio $R$ from polarization measurements and extracting $G_{Mp}$ from the measured cross section leads to  a renormalization of 2-3\% with respect to the Rosenbluth data, well inside the error bars \cite{Br03}. 
\item The inconsistency arises at the level of the slope of the $\epsilon$ dependence of the reduced cross section, which is  directly related to $G_{Ep}$, i.e., the derivative of the differential cross section, with respect to $\epsilon$. The difference of such slope, derived from the two methods above, appears particularly in the last and precise data \cite{Ar04}. One should note that the discrepancy appears in  the ratio $G_E/G_M$, whereas $G_M$, for example,  decreases more than one order of magnitude from $Q^2$=1 to 5 GeV$^2$. 

\end{itemize}

\begin{figure}
\begin{center}
\includegraphics[width=17cm]{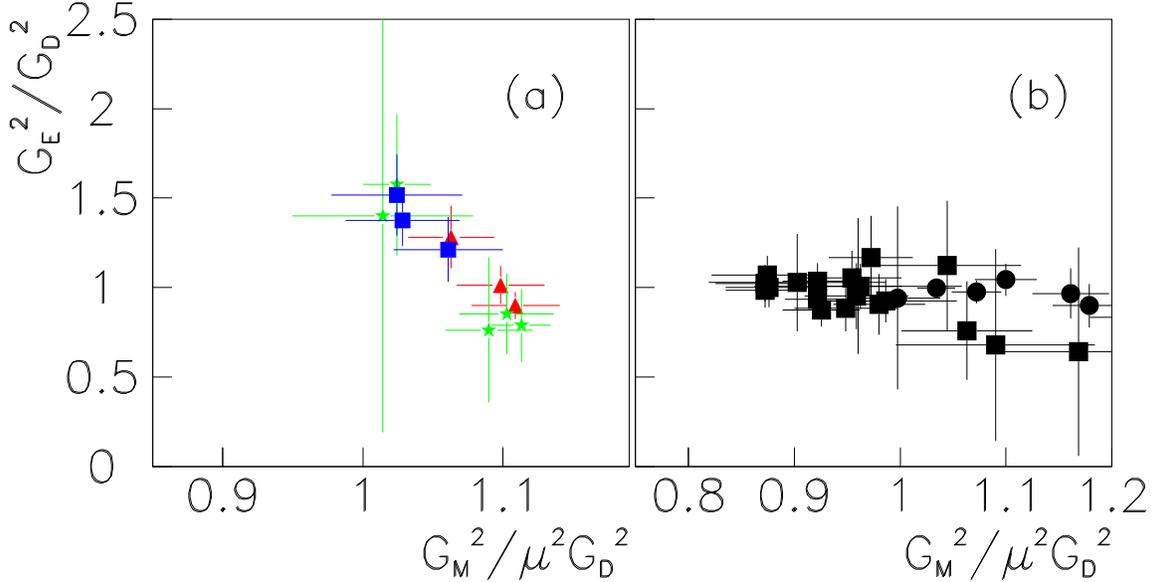}
\caption{\label{Fig:fig1} Dependence of  $G_E^2/G_D^2$ versus 
$G_M^2/\mu^2 G_D^2$:
(a) for $Q^2 \ge$ 2 GeV$^2$ from Refs. \protect\cite{Ar04} (triangles), 
\protect\cite{Ch04} (stars) and \protect\cite{Wa94} (squares); 
(b) for $Q^2 \le$ 2 GeV$^2$ from Refs. \protect\cite{Be71} (circles), 
and \protect\cite{Ja66} (squares). 
}
\end{center}
\end{figure}

The starting point of this work is the observation of a correlation, which appears in the published FFs data extracted with the Rosenbluth method: the larger is $G_E^2$, the smaller $G_M^2$. This is especially visible in the most recent and precise experiments, at large $Q^2$. The dependence of $G_E^2/G_D^2$ versus $G_M^2/\mu^2 G_D^2$  is shown in Fig. \ref{Fig:fig1}a for three recent data sets, at $Q^2\ge$ 2 GeV$^2$\cite{Ar04,Ch04,Wa94}. In Fig. \ref{Fig:fig1}b two data sets at low $Q^2$ ($Q^2 \le$ 2 GeV$^2$) are shown \cite{Be71,Ja66}. Whereas at low $Q^2$, $G_E^2/G_D^2$ seems constant and quite independent from $G_M^2/\mu^2  G_D^2$, at large $Q^2$ an evident correlation appears. 

Polarization data show also a linearity of the ratio $G_E/G_M$, but with an opposite trend. In this case, the ratio is measured directly, whereas according to the Rosenbluth method, one extracts two (independent) parameters from a linear fit. A correlation between the two parameters could be induced by the procedure itself or could be a physical effect and have a dynamical origin. In the last case, it should not depend on the experiment. In order to analyze this question in a quantitative way we have done a statistical study of the Rosenbluth data for several experiments. 

\begin{figure}
\begin{center}
\includegraphics[width=17cm]{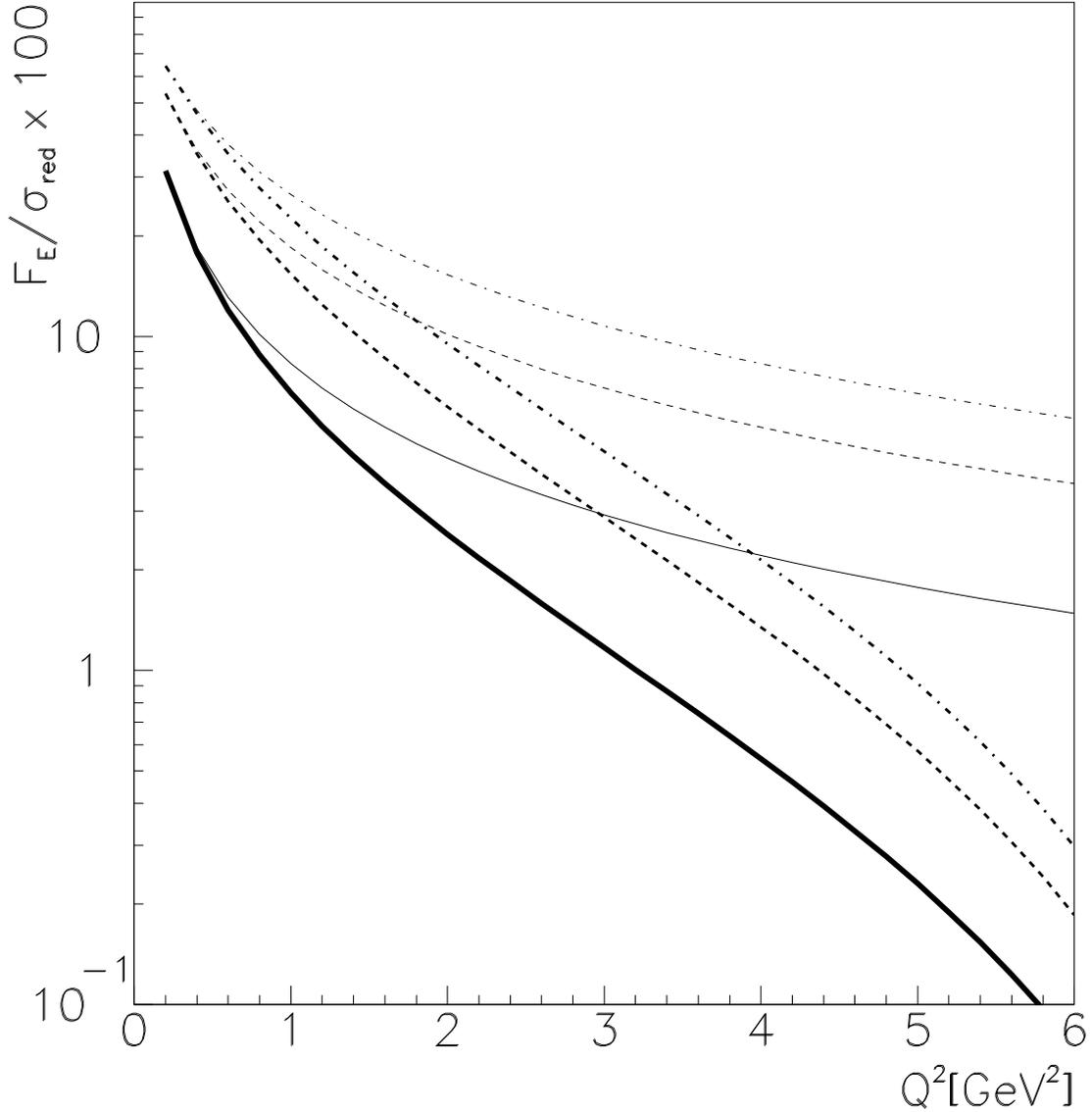}
\caption{\label{Fig:fig2} Contribution of the $G_E$ dependent term to the reduced cross section (in percent) for $\epsilon=0.2$ (solid line), $\epsilon=0.5$ (dashed line), $\epsilon=0.8$ (dash-dotted line), in the hypothesis of FF scaling (thin lines) or following Eq. (\protect\ref{eq:brash}) (thick lines).}
\end{center}
\end{figure}

First of all, at large $Q^2$, the contribution of the electric term to the cross section becomes very small, as the magnetic part is amplified by the kinematical factor $\tau $. This is illustrated in Fig. \ref{Fig:fig2}, where the ratio of the electric part ($F_E=\epsilon G_{Ep}^2$) to the reduced cross section is shown as a function of $Q^2$. The different curves correspond to different values of $\epsilon$, assuming FFs scaling or in the hypothesis of the linear dependence (\ref{eq:brash}). In the second case, one can see that, for example, for $\epsilon=0.2$, the electric contribution becomes lower than $3$\% starting from 2 GeV$^2$. This number should be compared with the uncertainty on the cross section measurement. When this contribution is larger or is of the same order, the sensitivity of the measurement to the electric term is lost and the extraction of $G_{Ep}$ becomes meaningless. 

Secondly, since the first measurements \cite{Ho62}, the electromagnetic probes are traditionally preferred to the hadronic beams, as the electromagnetic interaction is exactly calculable in QED, and one can safely extract the information from the hadronic vertex. However, one has to introduce the radiative corrections, which become very large as the momentum transfer squared, $Q^2$, increases. Radiative corrections were firstly calculated by Schwinger \cite{Shwinger} and are important for the discussion of the experimental determination of the differential cross section. 

The measured elastic cross section, is corrected by a global factor $C_R$, according to the prescription following \cite{Mo69}: 
\begin{equation}
\sigma_{red}= C_R\sigma_{red}^{meas}
\label{eq:sred}
\end{equation}
The factor $C_R$ {\it contains a large $\epsilon$ dependence}, and a smooth $Q^2$ dependence,  and it is common for the electric and magnetic part. At the largest $Q^2$ considered here, this factor can reach 30-40\%, getting larger when the resolution is higher. If one made a linear approximation for the uncorrected data, one might even find a negative slope starting from $Q^2\ge 3$ GeV$^2$ \cite{ETG}.

\begin{figure}
\begin{center}
\includegraphics[width=17cm]{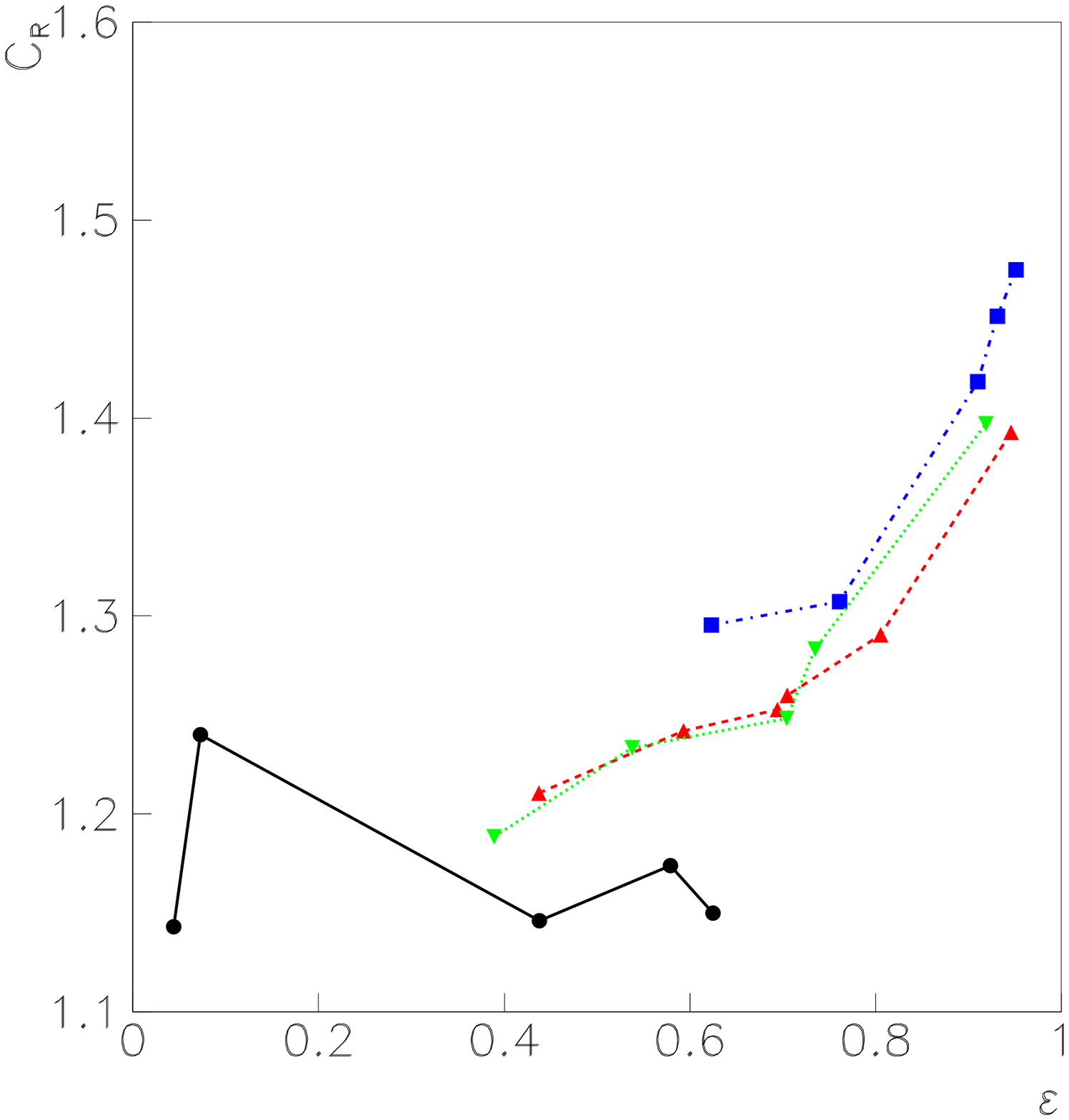}
\caption{\label{Fig:fig3} Radiative correction factor applied to the data at $Q^2$=3 GeV$^2$  (squares) from Ref. \protect\cite{Wa94}, at $Q^2$=4 GeV$^2$ (triangles) and 5 GeV$^2$ (inverted triangles) from Ref. \protect\cite{An94}, and at $Q^2$=0.32 GeV$^2$ from Ref. \protect\cite{Ja66} (circles). The lines are drawn to drive the eyes. }
\end{center}
\end{figure}

In Fig. \ref{Fig:fig3} we show the $C_R$ dependence on $\epsilon$, for different $Q^2$ and from different set of data. One can see that $C_R$ increases with $\epsilon$, arising very fast as $\epsilon\to 1$. It may be different in different experiments, because its calculation requires an integration on the experimental acceptance.

The Rosenbluth separation consists in a linear fit of the reduced cross section at fixed $Q^2$, where the two parameters are $G_E^2$ and $ G_M^2$.
Multiplying by a common factor, which depends strongly on $\epsilon$, the electric and magnetic term in (2) may induce a correlation between these two parameters. In order to test this hypothesis, we have built the error matrix for the Rosenbluth fits for different sets of data available in literature.  

\begin{figure}
\begin{center}
\includegraphics[width=17cm]{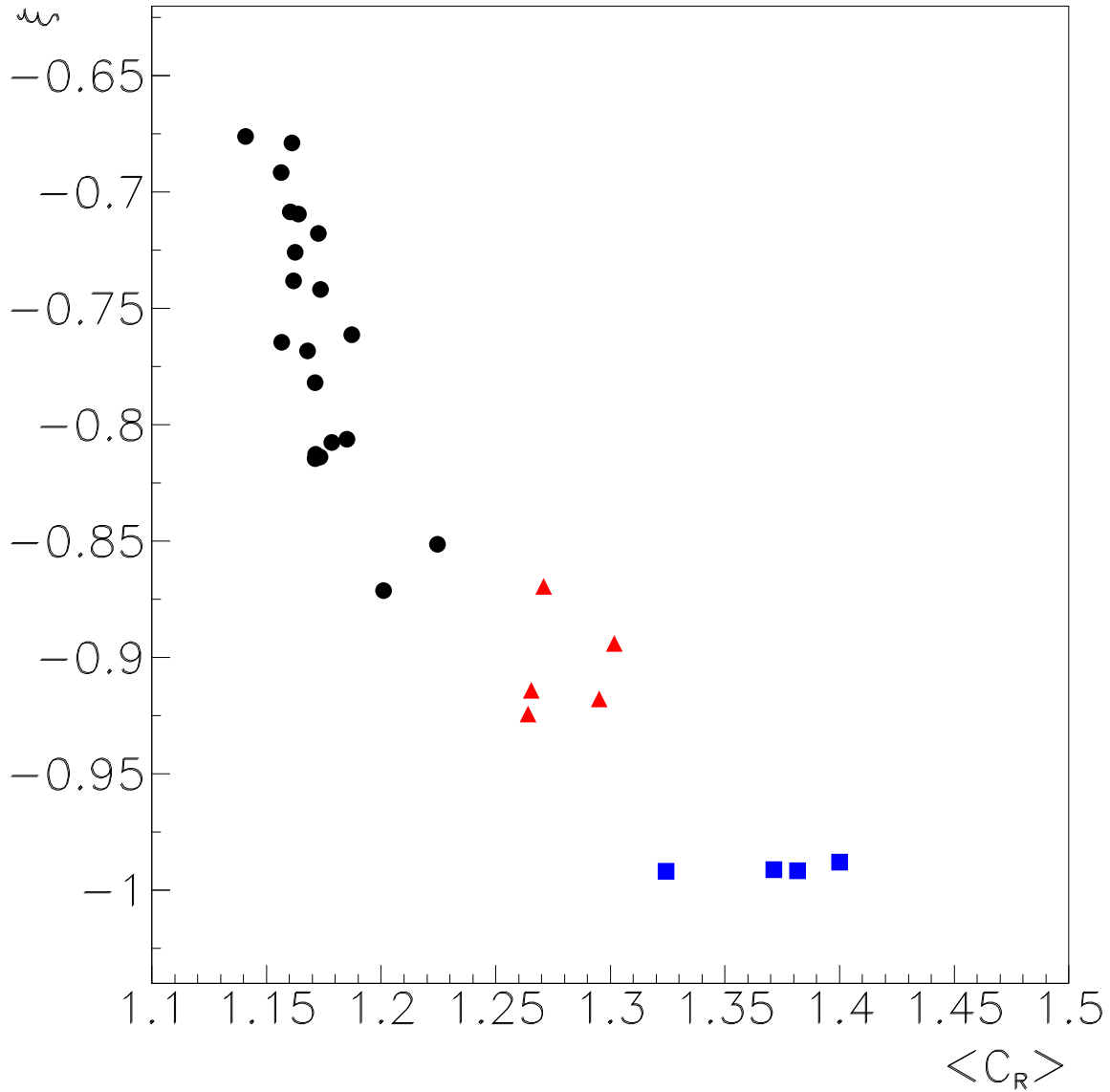}
\caption{\label{Fig:fig4} Correlation coefficient, $\xi$, as a function of the radiative correction factor $<C_R>$, averaged over $\epsilon$, for different sets of data: from Ref. \protect\cite{Ja66}  (circles), from Ref. \protect\cite{An94} (triangles) and from  Ref. \protect\cite{Wa94} (squares). }
\end{center}
\end{figure}

At fixed $Q^2$, the reduced cross section, normalized to $G_D^2$, has been parametrized by a linear $\epsilon$ dependence: 
$\sigma_{red}/G_D^2=a\epsilon +b$. The two parameters, $a$ and $b$, have been determined for each set of data as well as their errors $\sigma_a$, $\sigma_b$ and the covariance, $cov(a,b)$. The correlation coefficient $\xi$, is defined as
$\xi=cov(a,b)/\sigma_a\sigma_b$ and is shown in Fig. \ref{Fig:fig4} as a function of the average of the radiative correction factor $<C_R>$, weighted over $\epsilon$.
 
As the radiative corrections become larger, the correlation between the two parameters becomes also larger, reaching values near its maximum (in absolute value). Full correlation means that the two parameters are related through a constraint, i.e., it is possible to find a one-parameter description of the data. This does not necessarily means that the reduced cross section becomes flat, as a function of $\epsilon$, but that the slope is related to a kinematical effect, not to a dynamical one.

The data shown here correspond to three sets of experiments, where the necessary information on the radiative corrections is available. The correlation coefficient can be calculated for a larger number of data and one could plot the correlation as a function of $Q^2$. However, different experiments, at the same $Q^2$, have been done at different angles and energy, i.e., at different $\epsilon$, and the radiative corrections which enter in the determination of $a$ and $b$ are different. Such plot would be not easy to interprete.

At low $Q^2$ a correlation still exists, but it is smaller. For the data from Ref. \cite{Ja66} the radiative corrections are of the order of 15\%, seldom exceed 25\% and  correspond to  $\epsilon < 0.8$. This allows a more safe extraction of the FFs. 

Fig. \ref{Fig:fig4} shows that, for each $Q^2$, the extraction of FFs by a two parameters fit, may be biased by the $\epsilon$ dependence induced by the radiative corrections. Whatever the precision on the individual measurement is, the slope of the reduced cross section is not sensitive to $G_{Ep}$ at large $Q^2$, which, therefore, can not be extracted from the data.

The $Q^2$ dependence is therefore driven by $G_{Mp}$, which follows a dipole form. For each $Q^2$, a nonzero value of the ratio $G_{Ep}/G_{Mp}$  will lead to an apparent dipole dependence of $G_{Ep}$.

To summarize, we reanalyzed the Rosenbluth data with particular attention to the radiative corrections applied to the measured cross section, and we showed, from the (published) data themselves
that, at large $Q^2$ the contribution of $G_E$ to the cross section is so small that it can not be safely extracted. The method itself is biased, at large momentum transfer because the electric contribution to the measured cross section is in competition with the size of the $\epsilon$ dependent corrections. When plotting the reduced cross section as a function of $\epsilon$, one, indeed, sees a nonzero slope, but it is due to the $\epsilon$ dependence contained in the radiative corrections, and it is no more primarily related to the inner structure of the proton. 

Therefore, the Rosenbluth method can not be used to extract the nucleon FFs at large momentum transfer, due to an intrinsic limitation deriving from the large size of the radiative corrections, compared to the electric contribution to the differential cross section, and especially to their steep $\epsilon$ dependence. In other words, there is a type of systematic error which becomes dominant and has never been included in the data, preventing the extraction of $G_{Ep}$. We confirm the conclusion of a previous paper \cite{Re68}, which firstly suggested the polarization method for the determination of $G_{Ep}$, due to the increased sensitivity of the cross section to the magnetic term, at large $Q^2$: '{\it Thus, there exist a number of polarization experiments which are more effective for determining the proton charge form factor than is the measurement of the differential cross section for unpolarized particles}'.

This work was inspired by stimulating discussions with M. P. Rekalo. Thanks are due to J.L. Charvet, G.I. Gakh and  B. Tatischeff for useful suggestions and a careful reading of the manuscript.
{}

\end{document}